\documentclass[aps,prl,twocolumn,showpacs,amsmath,amssymb]{revtex4-1}
\usepackage{graphicx}
\usepackage{epstopdf}
\usepackage{dcolumn}
\usepackage{bm}

\usepackage[usenames,dvipsnames,table]{xcolor}
\usepackage[version=3]{mhchem} 
\usepackage{braket}
\usepackage{multirow}
\usepackage{soul}

\begin{document}
\title{A DMI guide to magnets micro-world}
\author{ V. V. Mazurenko$^1$, Y. O. Kvashnin$^{2}$, A. I. Lichtenstein$^{3,1}$, M. I. Katsnelson$^{4,1}$}
\affiliation{$^{1}$Theoretical Physics and Applied Mathematics Department,
Ural Federal University, Mira Street 19, 620002
Ekaterinburg, Russia \\
$^2$ Uppsala University, Department of Physics and Astronomy,
Division of Materials Theory, Box 516, SE-751 20 Uppsala, Sweden \\
$^{3}$I. Institut f{\"u}r Theoretische Physik, Universit{\"a}t
Hamburg, Jungiusstra{\ss}e 9, D-20355 Hamburg, Germany \\
$^{4}$ Radboud University Nijmegen, Institute for Molecules and Materials, Heyendaalseweg 135, NL-6525 AJ Nijmegen, The Netherlands }

\begin{abstract}
Dzyaloshinskii-Moriya interaction, DMI in short, represents an antisymmetric type of magnetic interactions that favour orthogonal orientation of spins and competes with Heisenberg exchange. Being introduced to explain weak ferromagnetism in antiferromagnets without an inversion center between magnetic atoms such an anisotropic interaction can be used to analyze other non-trivial magnetic structures of technological importance including spin spirals and skyrmions. Despite the fact that the corresponding DMI contribution to the magnetic energy of the system has a very compact form of the vector product of spins, the determination of DMI from first-principles electronic structure is a very challenging methodological and technical problem whose solution opens a door into the fascinating microscopic world of complex magnetic materials. In this paper we review a few such methods developed by us for calculating DMI and their applications to study the properties of real materials.
\end{abstract}

\maketitle

\section{Introduction}
In a seminal paper \cite{dzyaloshinskii} I.E.~Dzyaloshinskii has introduced a novel type of anisotropic magnetic interactions which are antisymmetric with respect to swapping the positions of two spins. 
This was done based on a purely phenomenological basis. 
Very soon, T.~Moriya \cite{moriya} suggested the first simplified microscopic explanation of these interactions, indirect exchange and spin-orbit coupling (SOC) being the key ingredients. 
The Hamiltonian governing these interaction can be written in the following form:
\begin{eqnarray}
\hat H_{\rm DMI} = \sum_{i,j} \mathbf{D}_{ij} [\hat {\bf S}_i \times \hat {\bf S}_j]
\label{DMHAMorig}
\end{eqnarray}
where ${\bf S}_i$ is the spin moment at the site $i$. 
Nowadays the parameter $\mathbf{D}_{ij}$, which is, by construction, is an axial vector, is known as Dzyaloshinskii-Moriya interaction (DMI). 
\\
\\
``\textit{Slow is the experience of all deep fountains: long have they to wait until they know what has fallen into their depths.}'' (F. Nietzsche). 
\\
\\
Whereas the first decades DMI were considered as more or less marginal subject in magnetism (with the only exception of the phenomenon of weak ferromagnetism) now they are the mainstream subject, of a great conceptual meaning and of a great practical importance~\cite{katsura,sergienko,mostovoy,bode,heide,rohart}. 
This only contribution would be sufficient to put the name of Igor Dzyaloshinskii among the main creators of modern physics of magnetism. 

We are very thankful to the organizers for their kind invitation to participate in the special issue dedicated to Dzyaloshinskii. In this short review we present our view on the fast growing field of DMI based mostly on our own experience of calculations and analysis of DMI parameters for specific magnetic materials. Starting with the simple model considerations close to the original formulation of Moriya we will discuss general computational tools used to calculate DMI for real magnetic materials, give a few examples of such calculations, and discuss some applications, first of all, related to magnetic skyrmions and other noncollinear magnetic configurations. 

\section{Methods for calculating the Dzyaloshinskii-Moriya interaction}
In this section, numerical approaches for calculating DMI are discussed.
We start with a microscopic theory by Moriya \cite{moriya} and show how it can be extended to analyze the dependence of DMI sign on the occupation of the $3d$ shell.  Then we will focus on a correlated band theory of the DMI, that is free from basic limitations of the superexchange theory and can be applied in a wide range of electronic Hamiltonian parameters corresponding to insulators and metals.
The last subsection of the methodological part is devoted to first-principles approaches based on the density functional theory.

\subsection{Microscopic theory of DMI}
The first microscopic theory of the antisymmetric anisotropic exchange interaction was developed by Moriya in 1960 and presented in Ref.\cite{moriya}.
It is based on the Anderson's idea on superexchange interaction \cite{AndersonPW} and formulated on the basis of the simplest electronic model accounting the on-site Coulomb interaction and the spin-orbit coupling on the level of the hopping integrals. Such an electronic model can be written in the following form
\begin{eqnarray}
\hat{\cal H}=\sum_{i j,\sigma\sigma'}t_{ij}^{\sigma\sigma'}\hat{a}_{i \sigma}^{+}\hat{a}_{j \sigma'}+\frac{1}{2}\sum_{i ,\sigma\sigma'}U \,\hat{a}_{i \sigma}^{+}\hat{a}_{i \sigma'}^{+}\hat{a}^{}_{i \sigma'}\hat{a}^{}_{i \sigma},
\label{Ham}
\end{eqnarray}
where $\hat{a}^{\dagger}_{i \sigma}$($a_{i \sigma}$) are the creation (annihilation) operators. $U$ is local Coulomb interaction,  $t_{ij}^{\sigma \sigma'}$ is the element of the spin-resolved hopping matrix. Formally, Eq.\ref{Ham} is nothing but the Hubbard model\cite{Hubbard, Gutzwiller, Kanamori} that was officially introduced three years later in 1963.
In the limit when the on-site Coulomb interaction is much larger than the hopping integrals
such a Hubbard model can be reduced to the spin model
\begin{eqnarray}
\hat {\mathcal{H}}^{spin} = \sum_{ij} J_{ij} \hat {\bf{S}}_{i} \hat {\bf{S}}_{j} + \sum_{ij} \mathbf{D}_{ij}  [\hat {\bf{S}}_{i} \times  \hat {\bf{S}}_{j}] + \sum\limits_{ij}\hat{\bf{S}}_i\overset{\leftrightarrow}{\Gamma}_{ij}\hat{\bf{S}}_j,
\label{spinham}
\end{eqnarray}
where $\hat {\boldsymbol{S}}$ is the spin operator, $J_{ij}$, $\mathbf{D}_{ij}$ and $\overset{\leftrightarrow}{\Gamma}_{ij}$ are the isotropic exchange interaction, antisymmetric anisotropic (Dzyaloshinskii-Moriya) and symmetric anisotropic interactions, respectively. The summation runs twice over all pairs.
In terms of the Hubbard Hamiltonian parameters the resulting expression for the DMI has the following form \cite{moriya, aharony}:
\begin{equation}
\label{DMMoriya}
\mathbf{D}_{ij}=-\frac{i}{2 U} [{\rm Tr_{\sigma}} \{ \hat t_{ji} \} {\rm Tr_{\sigma}} \{ \hat t_{ij}  \boldsymbol{\sigma} \}  -  {\rm Tr_{\sigma}} \{ \hat t_{ij} \} {\rm Tr_{\sigma}} \{ \hat t_{ji} \boldsymbol{\sigma} \}  ],
\end{equation}
where $\boldsymbol{\sigma}$ are the Pauli matrices.

Interestingly, the Moriya's microscopic theory was published in 1960, however, its first application to quantitative analysis of the magnetic properties of real materials was only done 30 years later by Coffey, Rice, and Zhang in Ref.\cite{Rice}. They have estimated ${\bf D}_{ij}$  for different phases of La$_2$CuO$_4$ and YBa$_2$Cu$_3$O$_6$ compounds. It was shown that peculiarities in the crystal structures of these systems result in different patterns of the DMI vectors and as the result different ground states with and without net magnetic moment can be realized.

An important feature of the one-band consideration of the DMI is that the Moriya's results were obtained by using an assumption of the constant $U$ value without orbital dependence as well as by neglecting the intra-atomic (Hund's) exchange contribution.
Further development of the microscopic theory of the antisymmetric anisotropic interaction was mainly related to its generalization to multi-orbital electronic Hamiltonians. As was shown in Ref.\cite{aharony} inter-orbital Coulomb and intra-atomic exchange interactions play an important role in formation of the DMI. However, their account on the model level results in complicated expressions for exchange interactions and necessity to define numerous electronic Hamiltonian parameters.

Another important peculiarity of the one-band consideration of the DMI was demonstrated in Ref.\cite{AharonyPRL}. It was shown that the resulting spin model, Eq.\ref{spinham} is characterized by a specific symmetry of the symmetric anisotropic exchange interaction tensor, $\overset{\leftrightarrow}{\Gamma}_{ij}$ whose principal axis coincides with DMI for each bond. It means that the state of a system with weak ferromagnetism is higher in energy than the pure (compensated) antiferromagnetic state. This result was confirmed by the authors of the same work \cite{AharonyPRL} on the level of the Hubbard model for which by means of the unitary transformation one can get exactly isotropic form of the electronic model. The latter gives purely isotropic Heisenberg model in the atomic limit.
Such a problem of the instability of the weakly ferromagnetic ground state of some antiferromagnets is solved by taking into account a multi-orbital nature of the transition metal oxides.

Despite of the above-mentioned and other limitations of the one-band approach for calculating magnetic interaction parameters, it provides a very simple and transparent way to analyze the properties of the interactions.  For instance, it can be used for analysis of the dependence of the DMI sign on the occupation of the $3d$ shell experimentally observed in the series of isostructural weak ferromagnets, MnCO$_3$, FeBO$_3$, CoCO$_3$, and NiCO$_3$ as it was done by us in Ref.\cite{carbonates}.
We consider the case of a transition metal oxide for which the crystal field splitting is much larger than the spin-orbit coupling, the latter can be treated as a perturbation. The corresponding expression for Dzyaloshinskii-Moriya interaction can be presented in the following form
\begin{eqnarray}
\mathbf{D}^{nn'}_{ij} = \frac{4i}{U} [b^{nn'}_{ij} \mathbf{C}^{n'n}_{ji} - \mathbf{C}^{nn'}_{ij}b^{n'n}_{ji}],
\end{eqnarray}
where $b^{nn'}_{ij}$ is the (unperturbed) hopping integral between $n^{th}$ ground orbital state of i$^{th}$ atom and $n'^{th}$ orbital state of j$^{th}$ atom, $\mathbf{C}^{nn'}_{ij}$ is the corresponding hopping renormalized by SOC and $U$ is the on-site Coulomb interaction.
Thus, $\mathbf{C}_{ji}^{n'n}$ is given by
\begin{eqnarray}
\mathbf{C}_{ji}^{n'n} = -\frac{\lambda}{2}[\frac{(\mathbf{L}^{m'n'}_{j})^*}{\epsilon_{j}^{m'}-\epsilon_{j}^{n'}}
b_{ji}^{m'n} +  \frac{\mathbf{L}^{mn}_{i}}{\epsilon_{i}^{m}-\epsilon_{i}^{n}}
b_{ji}^{n'm}],
\end{eqnarray}
where $\lambda$ is the spin-orbit coupling constant, $\mathbf{L}^{mn}_{i}$ is the matrix element of the orbital angular momentum between
the {\it m}th excited state and the {\it n}th ground state Wannier functions which are centered at {\it i}th ion,
while $\epsilon^{n}_{i}$ represents the energy of the {\it n}th Wannier orbital at the {\it i}th ion. \\

\begin{figure}
\includegraphics[width=0.8\columnwidth]{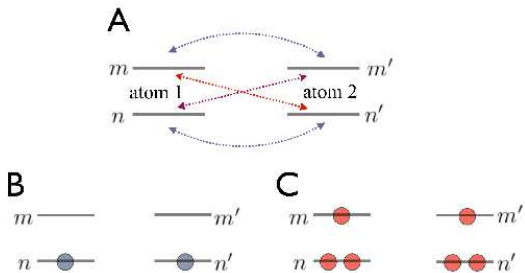}
\caption {(A) Minimal tight-binding model used for explaining the DMI sign change at variation of the occupation.
The horizontal lines represent the electron levels and hoppings are shown with arrows. (B and C) Two antiferromagnetic ground states corresponding to the $S=1/2$ case, obtained in the model for different orbital fillings: $N=2$ (left) and $N=6$ (right).}
\label{fig1theory}
\end{figure}

Tight-binding model we considered contains two atoms having non-degenerate ($n$ and $n'$) and high-energy ($m$ and $m'$) levels.
The schematic visualization of the model with the allowed hopping paths is presented in Fig.~\ref{fig1theory}.
In the simplest case one can assume that the same hopping integrals between high- ($m$ and $m'$) and low-energy ($n$ and $n'$) levels, $b_{12}^{m \, m'} = b_{12}^{n \, n'}$.
The hoppings between orbitals of different symmetry require more detail analysis, since they define the DMI in the system in question.
We assume that  the geometry of the model system is fixed, which means that hopping integrals do not change with variation of the occupation.

Our tight-binding model has two ground states with different occupations $N$ that correspond to the $S=1/2$ case: $N=2$ and $N=6$ (Fig.~\ref{fig1theory}).
In the case $N=2$, the ground state magnetic orbital is of symmetry $n$($n'$), while for $N=6$ it is $m$($m'$).
Another difference between these configurations is the different occupation of the excited states: they are empty and fully occupied for $N=2$ and $N=6$, respectively.

The difference between DMIs obtained for a system with two and six electrons is related to the difference between $\mathbf{C}_{21}^{n' \, n}$ and $ \mathbf{C}_{21}^{m' \, m}$,
\begin{eqnarray}
\mathbf{C}_{ji}^{n' \, n} = -\frac{\lambda}{2}[\frac{(\mathbf{L}^{m' \, n'}_{j})^*}{\epsilon_{j}^{m'}-\epsilon_{j}^{n'}}
b_{ji}^{m' \, n}
 +  \frac{\mathbf{L}^{m \, n}_{i}}{\epsilon_{i}^{m}-\epsilon_{i}^{n}}
b_{ji}^{n' \, m}]
\label{C1},
\end{eqnarray}

\begin{eqnarray}
\mathbf{C}_{ji}^{m' \, m} = -\frac{\lambda}{2}[\frac{(\mathbf{L}^{n' \, m'}_{j})^*}{\epsilon_{j}^{n'} - \epsilon_{j}^{m'}}
b_{ji}^{n' \, m}
+  \frac{\mathbf{L}^{n \, m}_{i}}{\epsilon_{i}^{n} - \epsilon_{i}^{m}}
b_{ji}^{m' \, n} ]
\label{C2}.
\end{eqnarray}

Using the relations for the orbital momenta matrix elements $\mathbf{L}^{m \, n} = - \mathbf{L}^{n \, m}$ and $\Delta E = \epsilon^{n}_{i} - \epsilon^{m}_{i}$ we rewrite Eqs.(\ref{C1}) - (\ref{C2}) in the following form:

\begin{eqnarray}
\mathbf{C}_{ji}^{n' \, n} = - \mathbf{C}_{ji}^{m' \, m}=  -\frac{\lambda \mathbf{L}^{mn}}{2 \Delta E}(
b_{ji}^{m' \, n} -
b_{ji}^{n' \, m}),
\end{eqnarray}

It means that   $\mathbf{D}^{nn'}_{ij}$ (for the system with two electrons) and $\mathbf{D}^{mm'}_{ij}$ (with six electrons) are of different signs.
Thus, on the level of Moriya's approach, the sign of the DMI depends on the occupation of the excited states.
Depending on the symmetry and occupation, each pair of $3d$ orbitals can result in positive or negative contribution to the total DMI between two atoms.
It should be noted that similar dependence of the DMI sign on the occupation of the $3d$ shell can be also found in some series of metallic systems.
In this sense interesting methodological results were obtained in Refs.\cite{DMIsign1,DMIsign2,DMIsign3}.

It is important to discuss the limits of the Moriya's theory of DMI from the point of view of its using to study real physical systems. In its original formulation it is limited to the systems with the spin state of $S = \frac{1}{2}$.  Real transition metal compounds and nanosystems are of multi-orbital nature. 
In this case, the main question is how to define the numerous hopping and Coulomb interaction parameters of the Hubbard model. 
In principle, one can use approximations of different types to define the parameters \cite{Moskvin1, Moskvin2} by using available experimental data.  Another approach is based on performing density functional theory (DFT) calculations and their parametrization using wannierization procedure developed in Refs.\cite{Wannier1,Wannier2} to construct the Wannier functions \cite{Wannier90}. Then, on this basis the electronic model parameters are calculated. The most accurate numerical scheme to estimate local ($U$) and non-local Coulomb interaction parameters taking screening effects into account is based on the constrained random phase approximation \cite{CRPA}.

The situation becomes even more complicated if one simulates a compound with a strong spin-orbit coupling. For this case effective numerical schemes based on the superexchange theory can be found in Refs.\cite{Solovyev1,Solovyev2}.  Another important limitation of the superexchange approach is that it is justified in the atomic limit, which is the case of the insulators. Its extension on the case of the metallic systems is not straightforward and requires account of the higher order terms in $t/U$ expansion. However, in all the cases the initial simplicity and transparency of the Moriya's consideration is lost. Below we will discuss the multi-band approaches for calculating DMI that are free from the perturbation theory applied for $t/U$ or spin-orbit coupling.


\subsection{Correlated band theory for DMI}
We start with correlated band theory of DMI developed by us in Ref.\cite{correlatedDMI}.
It is based on the consideration of the general Hamiltonian of interacting electrons in
a crystal:
\begin{eqnarray}
\hat H &=&  \sum_{12} c^{+}_{1} t_{12} c_{2} + \frac{1}{2} \sum_{1234} c^{+}_{1} c^{+}_{2} U_{1234} c_{3} c_{4},
\label{Ham1}
\end{eqnarray}
were $1=(i_1,m_1,\sigma_1)$ is the set of site $(i_1)$, orbital
$(m_1)$ and spin $(\sigma_1)$ quantum numbers and $t_{12}$ are
hopping integrals that contain the spin-orbit coupling. These transfer couplings can be found by the
Wannier-parameterization of the first-principle band structure with
the spin-orbit coupling.

We will take into account only the local Hubbard-like interactions,
keeping in $\hat H_{u}$ only terms with $i_1=i_2=i_3=i_4$. This
assumption corresponds to the DFT+U Hamiltonian \cite{LDAU} that
is also a starting point for the DFT+DMFT (Dynamical Mean-Field
Theory) \cite{LDA+DMFT,kotliar-DMFT, Anisimov}. It is crucially important
for the later consideration that the interaction term $\hat H_{u}$
is supposed to be rotationally invariant.

We start with a collinear magnetic configuration, for instance an antiferromagnetic state, which is close to the real ground
state (weak ferromagnet), but does not coincide with it due to the DMI. 
Let us re-define the DM Hamiltonian (Eq.\ref{DMHAMorig}) in a slightly different way:
\begin{eqnarray}
H_{\rm DMI}=\sum_{ij} {\bf D}^{\prime}_{ij} [{\bf e}_{i} \times {\bf e}_{j} ],
\label{DMHAM}
\end{eqnarray}
where ${\bf e}_{i}$ is a unit vector in the direction of the $i$-th
site magnetic moment and ${\bf D}^{\prime}_{ij}$ is the
Dzyaloshinskii-Moriya vector. We analyze the magnetic
configuration that is slightly deviated from the collinear state,
\begin{eqnarray}
{\bf e}_{i} = \eta_{i} {\bf e}_{0} + [ \delta \boldsymbol{\phi}_{i}
\times  \eta_{i} {\bf e}_{0} ], \label{delta}
\end{eqnarray}
where $\eta_{i}=\pm 1$, ${\bf e}_{0}$ is the unit vector along the
vector of antiferromagnetism, and $ \delta \boldsymbol{\phi}_{i}$ are
the vectors of small angular rotations.

Substituting Eq. (\ref{delta}) into Eq. (\ref{DMHAM}) one finds for
the variation of the magnetic energy:
\begin{eqnarray}
\delta E = \sum_{ij} {\bf D}^{\prime}_{ij} (\delta \boldsymbol{\phi}_{i} - \delta
\boldsymbol{\phi}_{j} ). \label{deltaE}
\end{eqnarray}

Now we should calculate the same variation for the microscopic
Hamiltonian (\ref{Ham1}). Similar to the procedure used in Ref.
\cite{katsnelson-epj} to derive exchange interactions for
the LDA+DMFT approach, we consider the effect of the local rotations
\begin{eqnarray}
\hat R_{i} = e^{i\delta \boldsymbol{\varphi}_{i} \hat {\bf J}_{i}},
\end{eqnarray}
on the total energy; here $\hat {\bf J}_{i} = \hat {\bf L}_{i} + \hat
{\bf S}_{i}$ is the total moment operator, $\hat {\bf L}_{i}$ and $\hat
{\bf S}_{i}$ are the orbital and spin moments, respectively.

The interaction part of the Hamiltonian $\hat{H}_{u}$ is
rotationally invariant and is not changed under this
transformation, opposite to the hopping part $\hat H_{t}$ :
\begin{eqnarray}
\delta \hat H_{t} = \sum_{ij} c^{+}_{i} ( \delta \hat R^{+}_{i} \hat t_{ij} + \hat t_{ij} \delta \hat R_{j} ) c_{j} \nonumber \\
= -i \sum_{ij} c^{+}_{i} ( \delta \boldsymbol{\phi}_{i} \hat {\bf J}_{i} \hat t_{ij} - \hat t_{ij} \hat {\bf J}_{j} \delta \boldsymbol{\phi}_{j} ) c_{j}.
\label{rotH}
\end{eqnarray}
Assuming that $\hat {\bf J}_{i} = \hat {\bf J}_{j} = \hat {\bf J}$ the change of the total energy takes the form
\begin{eqnarray}
\delta E = -\frac{i}{2} \sum_{ij}(\delta \boldsymbol{\phi}_{i} -
\delta \boldsymbol{\phi}_{j}) {\rm Tr}_{m, \sigma} \langle c^{+}_i [\hat
{\bf J} , \hat t_{ij}]_{+} c_{j} \rangle \nonumber \\
-\frac{i}{2} \sum_{ij}(\delta \boldsymbol{\phi}_{i} +
\delta \boldsymbol{\phi}_{j}) {\rm Tr}_{m, \sigma} \langle c^{+}_i [\hat
{\bf J} , \hat t_{ij}]_{-} c_{j} \rangle,
\label{rotE}
\end{eqnarray}
where ${\rm Tr}_{m, \sigma}$ is a trace over orbital ($m$) and spin
($\sigma$) quantum numbers.

The first term in the right-hand side of Eq. (\ref{rotE}) is
responsible for {\it relative} deviations of the magnetic moments
on sites $i$ and $j$ (DMI) whereas the second one is related with
the rotation of the magnetic axis as a whole (magnetic
anisotropy). Here we will focus on the DMI.

Comparing Eq. (\ref{rotE}) with Eq. (\ref{deltaE}) one finds
\begin{eqnarray}
\mathbf{D}^{\prime}_{ij}  = -\frac{i}{2} {\rm Tr}_{m,
\sigma} N_{ji} [\hat {\bf J}, \hat t_{ij}]_{+}, \label{vecD}
\end{eqnarray}
where $N_{ji} =\langle c^{+}_{i} c_{j} \rangle= -\frac{1}{\pi}
\int_{-\infty}^{E_{f}} {\rm Im} G_{ji} (E) dE$ is the inter-site occupation matrix
and $\hat G$ is the Green function of the system, $E_F$ is the Fermi energy.
The occupation matrix can be calculated by using a static (such as DFT+U \cite{LDAU}) or a dynamic mean-field approach (DFT+DMFT \cite{LDA+DMFT, kotliar-DMFT, Anisimov}).

Note that the occupation matrix is
calculated in the corresponding collinear states, which strictly
speaking can be done self-consistently only within constrained
calculations \cite{stocks}. Using the decomposition of the total
moment $\hat {\bf J}$ into orbital and spin moments, we have a
natural representation of the Dzyaloshinskii-Moriya vector
(\ref{vecD}) as a sum of the orbital and spin contributions which are related with the rotations
in orbital and spin space, respectively.

The resulting expression Eq.\ref{vecD} is of general nature and its spin part can be also derived in the case of the metallic systems as it was shown  \cite{Tatara}.

\subsection{DFT-based methods}

In this section we discuss mean-field approaches for calculating the DMI that are realized on the basis of the numerical methods of the density functional theory.
The net DMI can be assessed by calculating the DFT total energies for the two sets of spin spiral states having opposite helicities\cite{chshiev,zimmermann,sandratskii} or by using Berry phase theory\cite{Freimuth}.
In order to calculate the individual pair-wise DMI, one can employ the magnetic force theorem \cite{force}. 
According to this theorem, the variation of the total energy of the system due to a magnetic excitation can be expressed through the variation of the single-particle energy
\begin{eqnarray}
\delta E = - \int_{-\infty}^{E_{F}} \, d \epsilon \, \delta N (\epsilon),
\end{eqnarray}
here $N(\epsilon)$ is the integrated density of the electron state and $E_{F}$ is the Fermi energy.
Usually, the magnetic excitations related to a small rotation of the magnetic moments of the transition metal atoms from the collinear ground state are considered.
In this case the first and the second variations of the total energy written in the basis $|ilm\sigma\rangle$ (where $i$ denotes the site, $l$ the orbital quantum number, $m$- magnetic quantum number and
$\sigma$- spin index) are given by the following expressions
\begin{eqnarray}
\delta E=  - \frac{1}{\pi} \, \sum_{i} \int_{-\infty}^{E_{F}} d \epsilon \, {\rm Im} \, {\rm Tr}_{m,\sigma} \,  (\delta H_{i} \, G_{ii}) \label{firstvarE}
\end{eqnarray}
and
\begin{eqnarray}
\delta^{2} E=  - \frac{1}{\pi} \, \int_{-\infty}^{E_{F}} d \epsilon \, {\rm Im} \, {\rm Tr}_{m, \sigma} \, ( \sum_{i} \delta^{2} H_{i} \, G_{ii} \, \nonumber \\
+ \, \sum_{ij} \delta H_{i} \, G_{ij} \, \delta H_{j} \, G_{ji}).
\end{eqnarray}
Here $\delta H$ is the variation of the Hamiltonian, $G_{ii}$ and $G_{ij}$ are one-site and inter-site atomic Green's functions that can be calculated by using LDA+U approach.

Depending on the kind of the magnetic excitations we can define different parameters of the spin Hamiltonian for the atomic system. For instance, if the variation $\delta H_{i}$ is related to the rotation of the magnetic moments from the collinear ground state, then one can obtain the isotropic exchange interaction \cite{force}
\begin{eqnarray}
\label{J_Gr}
J_{ij} = -\frac{1}{4\pi S_i S_j} \int_{-\infty}^{E_{F}} d\epsilon \, {\rm Im}
{\rm Tr}_{m} (\Delta_{i} \,
G_{ij}^{\downarrow} \, \Delta_{j} \, G_{ji}^{\uparrow}),
\end{eqnarray}
where $\Delta_{i}$ is the magnetic splitting of the on-site potential and $S$ is the atomic spin.

In $3d$ systems, the spin-orbit coupling in itself can be also considered as a perturbation. \cite{Bruno,Solovyev}
One can consider a mixed perturbation scheme with respect to the rotation and spin-orbit coupling, which leads to the antisymmetric anisotropic DMI \cite{MnCuN}
\begin{eqnarray}
\label{Dz}
D^{z}_{ij} = - \frac{1}{8 \pi S_i S_j} \, {\rm Re}
\int_{-\infty}^{E_{F}} d \epsilon \, \sum_{k}
\nonumber \\
\times {\rm Tr}_{m} (\Delta_{i} G_{ik}^{\downarrow} H^{so}_{k \, \downarrow \downarrow} G_{kj}^{\downarrow} \Delta
_{j} G_{ji}^{\uparrow} -
\Delta_{i} G_{ik}^{\uparrow} H^{so}_{k \, \uparrow \uparrow} G_{kj}^{\uparrow} \Delta_{j} G_{ji}^{\downarrow}
\nonumber \\
+ \Delta_{i} G_{ij}^{\downarrow} \Delta_{j} G_{jk}^{\uparrow} H^{so}_{k \, \uparrow \uparrow} G_{ki}^{\uparrow} -
\Delta_{i} G_{ij}^{\uparrow} \Delta_{j} G_{jk}^{\downarrow} H^{so}_{k \, \downarrow \downarrow} G_{ki}^{\downarrow}).
\end{eqnarray}
Other components of the Dzyaloshinskii-Moriya vector for particular bond can be obtained from the $z$ ones by rotation of the coordinate system.
Similar expression for DMI was obtained by Solovyev \textit{et al}\cite{Solovyev-LaMnO3}.

The advantage of this perturbation-type consideration of the spin-orbit coupling is a methodological simplicity  of the DMI estimation. For this the required DFT calculations without spin-orbit coupling for different types of the collinear magnetic orderings can be routinely performed. Importantly, the formulation of the problem on the Green's functions language gives one opportunity to define the orbital contributions to the resulting DMI, which paves the way to a truly microscopic analysis of the anisotropic interactions. We have applied the developed method for calculating the DMI to give a microscopic explanation to the scanning tunneling microcopy experiments performed for chains of manganese atoms on CuN surface \cite{MnCuN}. Weak ferromagnetism due to the DMI between neighbouring manganese atoms was predicted. Another important  example is a first-principles study of the molecular nanomagnet Mn$_{12}$ for which most theoretical works on molecular magnets is mainly relied on the so-called rigid-spin model. Within such a model a complex system of interacting spins is replaced by just one big spin, with some magnetic anisotropy being introduced artificially. However, such a description is rather simplistic and largely ignores intermolecular interactions. Previously, it was predicted that the DMI plays a crucial role in the physics of molecular magnets  \cite{Harmon} and in particular magnetic tunneling effects in Mn$_{12}$ \cite{tunneling}.  In our work \cite{Mn12} we have demonstrated that the account of the inter-atomic anisotropic exchange interactions in Mn$_{12}$ gives opportunity to reproduce excitation energies observed in the inelastic neutron scattering experiments for this system.

In the case of the systems with strong spin-orbit coupling one could still use similar Green's function approach within the magnetic force theorem \cite{katsnelson}.
The expressions for DMI, which do not rely on the smallness of spin-orbit coupling constant have been derived independently by several groups\cite{udvardi,ebert,secchi,mankovsky,kvashnin2020}.
However, one of the main problems with large SOC systems is to stabilize a magnetic ordering of a collinear type in the first-principles calculations.
Due to a strong single-ion anisotropy the local magnetic moments pointing along local easy axes can form a stable non-collinear configuration. The attempt to implement a fixed moment procedure when the electronic structure calculations are performed with a pre-defined magnetic configuration can produce numerical instabilities in calculations results of different types. In this situation the using of the correlated band method for calculating DMI (Eq.\ref{vecD}) that taking spin and orbital contributions into account becomes well-motivated.

\section{Applications}

\subsection{Weak ferromagnetism in antiferromagnets}

\begin{figure}[!ht]
\includegraphics[width=0.5\columnwidth]{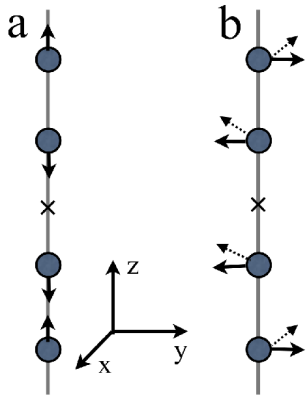}
\caption{Magnetic orderings with (right) and without (left) spin canting experimentally observed in Fe$_2$O$_3$. Cross denotes the inversion center.}
\label{Fe2O3}
\end{figure}

Discovery of the weak ferromagnetism in iron hematite, Fe$_2$O$_3$ (Ref.\cite{smith}) was the starting point for development of the DMI theory. More specifically, Fe$_2$O$_3$ is pure antiferromagnet with magnetic moments parallel to the trigonal axis $c$ at $T<$ 260 K (Fig.\ref{Fe2O3} left). In the temperature range between 260 and 950 K, the magnetic moments are in-plane and a small canting of the magnetic moment exists (Fig.\ref{Fe2O3} right). As the result of this canting there is net magnetic moment in the antiferromagnetic system. Such a difference between in-plane and out-of-plane magnetic structures is dictated by the symmetry of the system in question \cite{vonsovsky}. For the in-plane antiferromagnetic structure, for instance along $y$ axis there is symmetry operation related to the 180$^{\circ}$ rotation around $x$ axis. In general case such a symmetry does not require a pure antiferromagnetic ordering between neighbouring spins for which there is no inversion center. The corresponding rotation can be performed for canted magnetic moments.

A weak value of the net magnetic moment compared to the local magnetic moment, $M_{\rm net} / M_{\rm local} \sim 10^{-3}$ means that the force responsible for the canting is also very weak. At the microscopic level the absence of the inversion center between two spins results in a non-zero DMI in this spin pair. The weak ferromagnetism in Fe$_2$O$_3$ was explored with first-principles DFT calculations in Ref.\cite{Fe2O31} and with Green's function approach based on the magnetic force theorem in Ref.\cite{Fe2O32}.

\begin{figure}[!b]
\includegraphics[width=\columnwidth]{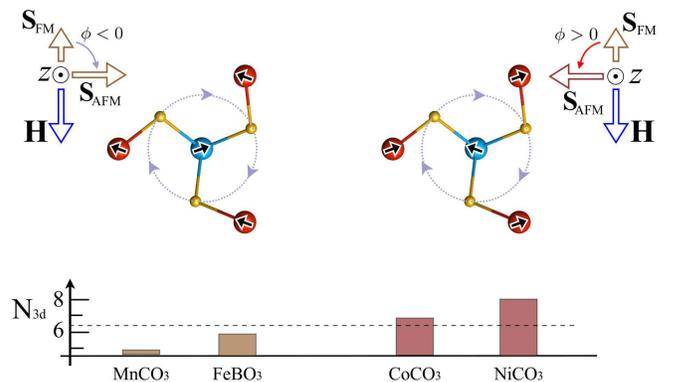}
\caption{Local atomic and magnetic orders in the weak ferromagnets.
 The ions of the two magnetic sublattices are represented by blue (site 1) and red (site 2) spheres, with black arrows denoting the direction of their spins.
Oxygen atoms between the two adjacent transition metal layers are represented as yellow spheres. The dotted circles highlight the twist of the oxygen layer.
The bottom panel shows the occupation of the 3d level of a magnetic ion.
The left and right panels show the two possible magnetic configurations which stabilize depending on the $3d$ occupation and, therefore, the sign of the DMI, for a net ferromagnetic moment pointing along the magnetic field $\mathbf{H}$.
 $\mathbf{S}_{AFM}$ denotes the direction of the antiferromagnetic spin structure. This figure is reproduced with permission from Ref.\cite{carbonates}}
 \label{FIG1}
\end{figure}

A more interesting situation concerning weak ferromagnetism in antiferromagnets is observed in transition metal oxides having calcite structure.
Previous magnetization measurements \cite{Kosterov2006, Petrov, Borovik1961,Kreines} have confirmed the existence of the non-compensate in-plane magnetization in several materials of this kind. However, the precise direction of the weak ferromagnetic moment with respect to the crystallographic axes and hence the "\textit{sign}" of DM interaction remained unknown. 
For the first time, this was unambiguously identified for FeBO$_3$ using resonant x-ray diffraction in Ref.~\cite{dmitrienko-natphys}.
We have performed \textit{ab initio} calculations and extracted the DM vectors using Eq.~\ref{vecD}. It was found that the theoretical calculations do not only reproduce the correct sign of DM vectors, but also give a very good estimate of the canting angle.

Next, we addressed the series of isostructural calcite oxides, namely: MnCO$_3$, FeBO$_3$, CoCO$_3$, NiCO$_3$. 
They all exhibit weak ferromagnetism and experiments based on the technique developed in Ref.~\cite{dmitrienko-natphys}, have revealed the change of canting angle sign across the series\cite{carbonates}. 
More specifically, the compounds MnCO$_3$ and FeBO$_3$ are characterized by the rotation sense which differs from that for the CoCO$_3$ and NiCO$_3$ systems. 
It is schematically shown in Fig.\ref{FIG1}. 
Taking into account that these compounds have the same crystal structure (and the same crystallographic chirality), such a sign change can be attributed to the difference in the occupation of the $3d$ shell.

To provide a theoretical support to these experiments in Ref.\cite{carbonates} we have performed first-principles calculations within local density approximation taking into account the on-site Coulomb interaction $U$ and spin-orbit coupling (DFT+$U$+SO). For these calculations, the initial magnetization directions were set to lie along $x$ direction.
Table~\ref{tab.abinitio} gives a comparison of the main theoretical and experimental results.  One can see that the resulting magnetic configuration is antiferromagnetic one characterized by a canting of the magnetic moments. Such a canted state is the lowest-energy state for all the systems under consideration.

One can see that the calculations reproduce the change of the DMI sign through the series of studied compounds, observed experimentally.  Another important result is that first-principles calculations have revealed the increase of the canting angle absolute value as the 3d shell occupation increases, which also agrees with experimental data. According to the calculated occupation numbers the chemical bonding in all four systems has more covalent rather than ionic character, as indicated
by the deviation of the number of the $3d$ electrons from the pure ionic values, and magnetization of the oxygen atoms.

\begin{table}
\caption [Bset]{Occupation of the 3d shell and magnetic moments (in $\mu_B$) obtained from the first-principles calculations \cite{carbonates} for the weak ferromagnets we consider.
The first $3d$ metal atom is located at the origin,
while the second one is at (1/3, 2/3, 1/6) in the hexagonal settings.
The canting angle $\phi$ is defined as $\arctan(M_y/M_x)$ with the appropriate sign. The corresponding experimental estimates of $\phi$ with the references are presented in the brackets. The table was adopted from Ref.\cite{carbonates}.}
\begin {tabular}{ccccccccccc}
\hline
\hline
Compound    & $M_x$ & $M_y$ & $M_z$  &
$\phi$ (deg)  \\
\hline
                 & -4.503 & -0.004 & 0      &     -0.05    \\
MnCO$_3$     &  4.503 & -0.004 & 0      & (-0.04 \cite{Kosterov2006})\\
\hline
                & -4.138 & -0.057 & 0      &   -0.8      \\
FeBO$_3$     &  4.138 & -0.057 & 0      & (-0.9 \cite{Petrov}) \\
\hline
                 &  3.314 & -0.274 & -0.023 &     4.7   \\
CoCO$_3$     & -3.314 & -0.274 &  0.023 &  (4.9 \cite{Borovik1961,Kreines}) \\
\hline
                 &  1.792 & -0.233 & 0      &   7.4     \\
NiCO$_3$     & -1.792 & -0.233 & 0      &  (10.8 \cite{Kreines})   \\

\hline
\hline
\end{tabular}
\label{tab.abinitio}
\end{table}

Thus, the sign, symmetry and magnitude of DMI in MnCO$_3$, FeBO$_3$, CoCO$_3$ and NiCO$_3$ can be fully reproduced by means of the all-electron  DFT+$U$+SO calculations. However, the full calculation does not provide a truly microscopic understanding of the DMI sign change phenomena. 
For that a minimal tight-binding model based on the Moriya's theory as described in the methodological part of this paper becomes extremely useful.

\subsection{Magnetic skyrmions}
Investigation of skyrmions is a widely studied topic in the modern material science. 
From the very beginning, magnetic skyrmions in condensed matter physics have been introduced as classical topological spin structures \cite{Bogdanov}. Skyrmions usually appear as the result of a competition between different types of magnetic interactions. The most well-known case leading to the formation of a skyrmionic spin texture is the presence of the Dzyaloshinskii-Moriya interaction in the system \cite{Muhlbauer, Neubauer, Munzer, Yu, Nagaosa, Janson}. The latter competes with the exchange interaction and is responsible for the canting of spins from the collinear ordering. Since the DMI is typically much weaker than the exchange interaction, the characteristic size of resulting skyrmions is relatively large and may reach several hundred angstroms. This allows to describe these topological spin structures in the framework of classical micromagnetic models, where the magnetization of the lattice is treated as a continuous classical vector field. This classical theoretical description is used even for very compact skyrmion structures of 1-2 nm that have been experimentally found in surface nanostructures \cite{Blugel_skyrm, Wiesendanger}.

Previous experimental and theoretical studies on the topologically-protected magnetic skyrmion excitations were fully focused on the transition metal crystals  and nanosystems. This seems to be natural, since these materials are characterised by well-localised magnetic moments originated from the partially-filled 3d states and magnetic anisotropy that facilitates an experimental detection of the distinct magnetic textures. In works \cite{C2F, Si,SiC} a new class of materials, surface nanostructures with $sp$ element revealing skyrmion excitations at experimentally achievable magnetic fields and temperatures was introduced. The non-trivial result, that such $sp$-electron systems are, in principle, characterized by a magnetic state,  was experimentally confirmed in Refs.\cite{Glass,Li}.  Another important experimental result is that  the magnetic state is long-range, it is not localized at a specific atom of the system as in transition-metal compounds. For instance, STM experiments \cite{Hgraphene} on single or dimer hydrogen adatoms deposited on graphene have shown that the spin-polarized state extends over several nanometers away from the hydrogen atoms. It provides a direct exchange coupling between magnetic moments at long distances.

\begin{figure}[!h]
\includegraphics[width=1.0\columnwidth,angle=0]{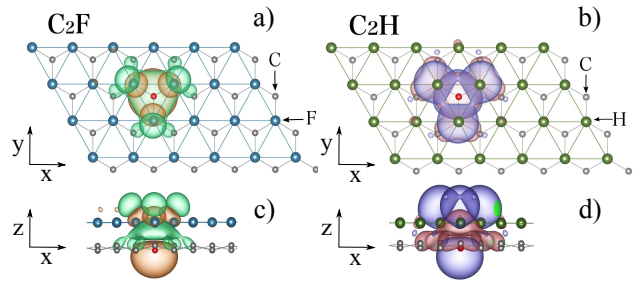}
\caption{(Color online) Wannier functions describing the band at the Fermi level in C$_{2}$F (a,c) and C$_2$H (b,d). Red sphere denotes the center of the Wannier orbital. This figure is reproduced with permission from Ref.\cite{C2F}.}
\label{ris:wannier_functions}
\end{figure}

Our first-principles calculations \cite{C2F, Si,SiC} have confirmed a long-range character of the magnetic states in graphene derivatives C$_2$H and C$_2$F as well as in surface nanostructures Si(111):\{C, Si, Sn, Pb\} and in Sn on  SiC(0001). Examples of the corresponding Wannier functions in semifluorinated and semihydrogenated graphene presented in Fig.\ref{ris:wannier_functions} demonstrate that substantial amount of the electron density is concentrated in the interstitial region. In all the cases we have found that the values of the calculated hopping integrals are much smaller than that of the Coulomb interactions, which gives us opportunity to construct a Heisenberg-type Hamiltonian for the localized spins $S=1/2$ within the superexchange theory

The calculations of the inter-site exchange interaction have confirmed a strong ferromagnetic contribution from the direct exchange interaction that can fully compensate the so-called antiferromagnetic Anderson's superexchange.
Another important contribution to the magnetic energy of the system comes from anisotropic Dzyaloshinskii-Moriya interaction. Importantly,  there is a conceptual difference in the origin of magnetic anisotropy in materials with $sp$ electrons that we consider and transition-metal compounds with localized $d$ electrons \cite{Wenzel,Torun}.  While the magnetic anisotropy in $3d$, $4d$ and $5d$ systems originates from the spin-orbit coupling of individual metallic atoms, it is not the case of materials with $sp$ electrons. Due to a strong delocalization of the magnetic moments  the magnetic anisotropy in these $sp$ materials is a collective multi-atomic effect. The resulting DMI vectors between Wannier functions in the systems in question were calculated by using the original Moriya's theory, Eq.\ref{DMMoriya}. In Ref.\cite{C2F} we have performed a detail symmetry analysis of the resulting magnetic model by using the Moriya's rules derived in the original work \cite{moriya}. It was shown that the resulting symmetry of the effective model of C$_2$F system is $C_{3}$, which allows an alternation of the $z$ component of the anisotropic exchange parameters within the coordination sphere under $C_{3}$ rotations.

\begin{figure}[!h]
\includegraphics[width=1.0\columnwidth,angle=0]{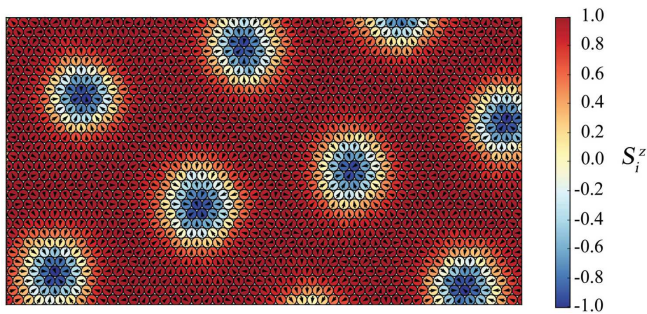}
\caption{Skyrmionic magnetic structure obtained from the Monte Carlo simulations for C$_2$F system. This figure is adopted from Ref.\cite{C2F}.}
\label{skyrmion}
\end{figure}

The constructed spin models for graphene derivatives, Si(111):\{C, Si, Sn, Pb\} and Sn monolayer on SiC(0001) surface
 were solved by means of the Monte Carlo methods, which gives us opportunity to define the magnetic phases of these materials depending on the external magnetic field and temperature. It was found that one can stabilize skyrmionic solutions in the case of the semifluorinated  graphene and nanosystems with heavy adatoms Sn/Si(111), Pb/Si(111) and Sn/SiC(0001). The key quantity here allowing the formation of the topologically protected magnetic structures is the anisotropic Dzyaloshinskii-Moriya interaction.

\section{Perspectives}
\begin{figure}[!h]
\includegraphics[width=1.0\columnwidth,angle=0]{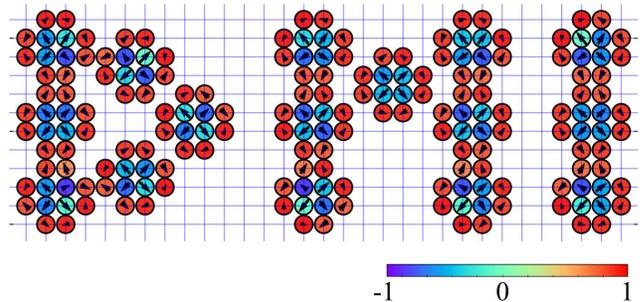}
\caption{Abbreviation of Dzyaloshinskii-Moriya interaction written with nanoskyrmions. This is a result of Monte Carlo simulations of the Heisenberg-exchange-free model on the square lattice with non-regular site occupation. Arrows and colors depict the in-plane and out-of-plane spin projections, respectively.}
\label{nanoskyrmion}
\end{figure}

Despite there were several decades of intensive investigations on DMI and related phenomena we think that this is still young and very promising research field within which one could focus on the following directions for future investigations. From the very beginning DMI is considered as a representative of one of the smallest energy scales in the magnetic Hamiltonians of the strongly correlated systems, $|{\bf D}_{ij}| < <J_{ij}$. It is due to the DMI always contains additional relativistic small parameter, the ratio of electron velocity in atoms to the velocity of light. However, such a dominance of the Heisenberg exchange interaction can be overcome by different means. For instance, manipulation of magnetic interactions via a strong periodic in time electromagnetic field \cite{Stepanov1} (''Floquet engineering'') suggests that using real nanosystems and real values of the laser fields one can reach the regime when the Heisenberg exchange $J_{ij}$ is arbitrarily small, or even equal to zero, whereas the Dzyaloshinskii-Moriya parameter ${\bf D}_{ij}$ remains constant. As an interesting example of such a situation, a new class of two-dimensional Heisenberg-exchange-free materials where a completely new type of skyrmions (Fig.\ref{nanoskyrmion}) that emerge as the result of the competition between the DMI and uniform magnetic field has been introduced \cite{Stepanov2}.

Another fascinating research field is related to a quantum skyrmions that in contrast to the classical counterpart are practically unexplored. The main methodological problem here is how to characterize the topology of quantum system with a three-dimensional magnetic structure when the orientation of a spin is ill-defined. One of the possible solutions was recently proposed by some of us in Ref.\cite{quantumskyrmion} where scalar chirality operator was introduced to define a quantum analog of the topological charge.

The other direction we consider to be very attractive is DMI applications in quantum computing. Existence of the anisotropic exchange interactions between qubits can be used for preparing highly entangled quantum states that play an important role in quantum information processing \cite{Grover, Shor}. 
The future of DMI looks bright and promising.

\section{Acknowledgements}
The work of A.I.L. and  M.I.K. is supported by European Research Council via Synergy Grant 854843 - FASTCORR.
Y.O.K. acknowledges the financial support from the Swedish Research Council (VR) under the project No. 2019-03569.
The work of V.V.M. was supported by Act 211 Government of the Russian Federation, contract 02.A03.21.0006.

\end{document}